\documentclass[20pt]{article}

\usepackage{PRIMEarxiv}

\usepackage[utf8]{inputenc} 
\usepackage[T1]{fontenc}    
\usepackage{hyperref}       
\usepackage{url}            
\usepackage{booktabs}       
\usepackage{amsfonts}       
\usepackage{nicefrac}       
\usepackage{microtype}      
\usepackage{lipsum}
\usepackage{fancyhdr}       
\usepackage{graphicx}       
\graphicspath{{media/}}     
\usepackage{xcolor}
\hypersetup{colorlinks=true, linkcolor=blue, filecolor=black, urlcolor=blue, citecolor=green}
\usepackage[acronym,toc]{glossaries}
\usepackage[intoc]{nomencl}
\usepackage{appendix}

\usepackage{amsfonts, amsmath, amsthm, amssymb}
\usepackage{newtxtext,newtxmath}
\title{Classical Soft Graviton Theorem to Memory Effect and Violation of Peeling}

\author{Raikhik Das \\
  Department of Physics \\
  Indian Institute of Science Education and Research, Pune \\
  Dr. Homi Bhabha Rd, Pashan, Pune, Maharashtra 411008, India\\
  \texttt{raikhikd@gmail.com} \\
}

\numberwithin{equation}{subsection}

\begin{document}
\maketitle

\begin{abstract}
It has been known for some time that the asymptotic structure of spacetime and soft theorems are closely related. To study the structure of future null infinity, studying the classical soft graviton theorem is often quite helpful. The memory effect at the future null infinity can be demonstrated from the leading behavior of gravitational radiation low-frequency. However, the memory effect is not the only information we can get from soft gravitational radiation. This paper demonstrates how the classical soft graviton theorem enlightens us about the memory effect and the differential structure of the future null infinity.
\end{abstract}


\section{Introduction}

\subsection{Peeling and Partial Peeling Property}
Bondi, van der Burg, Metzner \cite{Bondi} and Sachs \cite{Sachs} showed that in the presence of spatially compact sources, the future null infinity 
$(\mathcal{I}^+)$ in the asymptotically flat (AF) spacetimes is $C^\infty$. Penrose \cite{Penrose1, Penrose2} proposed a notion of asymptotic flatness suggesting the "Asymptotic Simplicity" condition according to which the future null infinity of AF spacetimes is $C^\infty$. Later, Geroch \cite{Geroch} also provided a notion for asymptotic flatness. Ashtekar and Hansen \cite{Ashtekar1, Ashtekar2} combined those two conditions. The asymptotic simplicity condition directly led to Penrose's peeling property: $C_{\mu \nu \rho \sigma} = \mathcal{O}(\Omega)$ at $\mathcal{I}^+$, where $C_{\mu \nu \rho \sigma}$ is the Weyl tensor, and $\Omega$ is the conformal factor. This model of AF spacetimes at $\mathcal{I}$ has provided a robust formalism for understanding gravitational radiation and its effect at $\mathcal{I}^+$.

However, several studies have provided evidence that the peeling property at $\mathcal{I}^+$ may not hold in AF spacetimes. Scalar, electromagnetic, and gravitational field perturbation of Schwarzschild background violates the peeling property at $\mathcal{I}$ \cite{Bardeen}. The gravitational radiation from a collapsing gas cloud with Newtonian limit violates the peeling property at $\mathcal{I}$. Christodoulou \cite{Christodoulou} has demonstrated that in the presence of $N$ massive particles at far past with the Newtonian limit on Minkowski background, peeling is violated at $\mathcal{I}^+$ as $u \to -\infty$. In Christodoulou's analysis, the $\log{r}/r$ term appears in the metric, where $r$ is the radial distance. This $\log{r}/r$ term in the metric makes the future null infinity $C^2$ as $u \to -\infty$. Hence, the peeling is violated. Winicour \cite{Winicour} also talks about spacetimes that are not asymptotically simple and follow the partial peeling property ($C_{\mu \nu \rho \sigma} = \mathcal{O}(\Omega \log{\Omega})$) rather than the peeling property. 

\subsection{Soft Theorem and Asymptotic Structure of Spacetime}
A deeper dive into soft theorems and the asymptotic structure of spacetimes has revealed that the apparently two different topics are equivalent \cite{Sen, Strominger2016}. Hence, studying one topic can enlighten us about the other. Peeling at the future null infinity can also be studied from the point of view of classical soft graviton theorems \cite{Sen, Sahoo2022, Sen_Laddha}. In Ashoke Sen et al.'s work \cite{Sen}, one can see that due to particles at far past, terms proportional to $\log{\omega}$ appear in the analysis of infrared gravitational radiation. Sen et al. use perturbed metric instead of pure flat metric as the background to calculate the subleading terms of soft radiation. Upon accounting for the non-linear effects, Sen et al.'s analysis demonstrates the existence of $\log{\omega}$ terms in the case of massive particles. This paper shows how the $\log{\omega}$ terms, in the case of soft radiation, lead to the violation of peeling at $\mathcal{I}^+$. This analysis implies the violation of peeling at $\mathcal{I}^+$ as $u \to + \infty$ and $u \to - \infty$.

\subsection{Organization}

Though all the analysis has been done for $4$-D in this article, it can be generalized to $D>4$ easily. In recent works and previously, the gravitational memory effect has been discussed a lot more than the violation of peeling at the future null infinity. Christodoulou \cite{Christodoulou} shed light on the violation of peeling, but it has gone vastly unnoticed in the physics community for the most part. Damour demonstrated a violation of peeling on Schwarzschildian background. Winicour \cite{Winicour} studied the asymptotic structure of LAF spacetime. Winicour's study involved calculating the metric components starting from a weaker peeling condition on the celestial sphere. Geiller and Zwikiel \cite{Geiller} have extensively studied the structure of LAF spacetimes. But in both studies about the structure of LAF spacetimes, the $\log$ terms in the metric are introduced by hand. This article justifies the existence of $\log$ terms in their work and demonstrates how the classical soft graviton theorem can play a significant role in doing so. The existence of $\log$ terms in the metric at null infinity opens a vast scope of studies about the asymptotic symmetries of LAF spacetimes and the charges corresponding to those symmetries instead of asymptotically simple spacetimes. It also opens the horizons for studying Celestial Holography corresponding to LAF spacetimes.

The sec.[\ref{Sec_Violation of Peeling}] briefly discusses Christodoulou's analysis of how peeling can be violated and how linearized perturbation to the background metric can be helpful in the prediction of violation of peeling. Then, brief reviews of linearized gravitation radiation and the classical soft graviton theorem are presented in sec.[\ref{Sec_Grav_Rad}] and sec.[\ref{Sec_CSGT}], respectively. Sec.[\ref{Sec_Behav}] describes the behaviour of the spacetime at $\mathcal{I^+_\pm}$, leading from sec.[\ref{Sec_CSGT}]. In sec.[\ref{Sec_Eff_Sft_Grav}], it is described how the soft gravitational radiation influences the structure of $\mathcal{I}^+$. Sec.[\ref{Sec_Eff_Sft_Grav}] discusses the memory effects and violation of peeling as consequences of the different terms in the spectrum of low-frequency gravitational radiation. And finally, sec.[\ref{Sec_Rslt_Disc}] summarizes and discusses the results of this article. 

\section{Violation of Peeling}\label{Sec_Violation of Peeling}

For peeling to be violated at $\mathcal{I}^+$, $\mathcal{I}^+$ has to have differentiability less than $3$. It is possible for the differentiability of $\mathcal{I}^+$ to be less than $3$ if the metric at $\mathcal{I}^+$ has non-artifact $\log{r}$ terms in them. The non-artifact $\log{r}$ terms in the metric will come due to the physical system rather than the coordinate choice.

Following Christodoulou's work \cite{Christodoulou}, one can define: 
\begin{equation}
    \Xi^\pm = \lim_{u \to \pm\infty} u^2 N_{\mu\nu}(x),
    \label{Christ_Anls_1}
\end{equation}
where $u=t-r$ and $N_{\mu\nu}(x)$ is the news tensor. It can be shown that 
\begin{equation}
    \lim_{r \to \infty} \partial_u (r^4 \beta) = \frac{\mathcal{D}^{(3)} \Xi^\pm}{|u|},
    \label{Christ_Anls_2}
\end{equation}
where $\mathcal{D}^{(3)}$ represents a third order differential operator on unit sphere $\mathbf{S}^2$. If we integrate eq.\eqref{Christ_Anls_2}, we get
\begin{equation}
    \lim_{u \to \pm\infty}\left((r^4 \beta)(u_2(t),t) - (r^4 \beta)(u_1(t),0)\right) \sim \int_{u_1(t)}^{u_2(t)} \frac{\mathcal{D}^{(3)} \Xi^\pm}{|u|} du\sim \left(\log{r} - \log{|u|}\right) \:\mathcal{D}^{(3)} \Xi^\pm.
    \label{Christ_Anls_3}
\end{equation}

As past stationarity is assumed in Christodoulou's analysis, one can set $(r^4 \beta)(u_1(t),0)=0$. Therefore, 
\begin{equation}
    \lim_{\mathcal{I}^+, u \to \pm\infty}\beta= B^* \frac{\log{r} - \log{|u|}}{r^4},
    \label{Christ_Anls_4}
\end{equation}
where $B^*$ depends on the quadrupole distribution. The expression in eq.\eqref{Christ_Anls_4} points towards the following asymptotic form of the metric at $\mathcal{I}^+_\pm$:
\begin{equation}
    g_{AB} = r^2 \, q_{AB} + r \, c_{AB} + d_{AB} + \frac{j_{AB} \ln{r}}{r} + \mathcal{O}\left(\frac{1}{r}\right) \: , \text{ in Bondi gauge.}
    \label{Christ_Anls_5}
\end{equation}

In eq.\eqref{Christ_Anls_5}, $d_{AB}$ and $j_{AB}$ depend on $c_{AB}$ following Bondi gauge. $q_{AB}$ is the two sphere metric: $d\Omega^2=d\theta^2+\sin^2{\theta}\,d\phi^2$.

The $\log{r}$ term appearing in eq.\eqref{Christ_Anls_4} and eq.\eqref{Christ_Anls_5} is not an artifact term. Instead, it appears due to the system considered. And due to the $\log$ term, the spacetime becomes $C^2$. Hence, if $\mathcal{D}^{(3)}\Xi^\pm$ is non-zero and finite, peeling will be violated at $\mathcal{I}^+_\pm$. So, to investigate the violation of peeling at $\mathcal{I}^+_\pm$, one can just check if $\mathcal{D}^{(3)}\Xi^\pm = \lim_{u \to \pm\infty} u^2 \partial_u \mathcal{D}^{(3)} c_{AB}$ is finite and non-zero. This motivates the study of the leading order perturbation to the background metric. With this motivation, linearized perturbation to the background metric will be studied in sec.[\ref{Sec_Grav_Rad}].

\section{Linearized Gravitational Radiation}\label{Sec_Grav_Rad}

In light of our interest in the linearized perturbation to the metric, as stated in sec.[\ref{Sec_Violation of Peeling}], linearized perturbation to the Minkowski metric is briefly reviewed in this section. Let's assume that we have a metric $g_{\mu\nu}$ such that
\begin{equation}
    g_{\mu\nu} = \eta_{\mu\nu} + 2 h_{\mu\nu},
    \label{GW_1}
\end{equation}
where $\eta_{\mu\nu}$ is the Minkowski metric and $h_{\mu\nu}$ part represents the perturbation to $\eta_{\mu\nu}$.\\

Now, let's define:
\begin{equation}
    e_{\mu\nu} = h_{\mu\nu} - \frac{1}{2} \eta_{\mu\nu} \eta^{\alpha\beta} h_{\alpha\beta}.
    \label{GW_2}
\end{equation}

Therefore,
\begin{equation}
     h_{\mu\nu} = e_{\mu\nu} - \frac{1}{2} \eta_{\mu\nu} \eta^{\alpha\beta} e_{\alpha\beta}.
    \label{GW_3}
\end{equation}

In de-Donder coordinates or harmonic coordinates, Einstein's equations of general relativity for $g_{\mu\nu}$ reduces to:
\begin{equation}
    \square e_{\mu\nu} = -8 \pi G T_{\mu\nu},
    \label{GW_4}
\end{equation}
where $\square \equiv \eta_{\alpha\beta \partial_\alpha \partial_\beta}$, $G$ is Newton's universal constant and $T_{\mu\nu}$ is the stress-energy tensor.\\

For convenience, one can work in units for which $8 \pi G = 1$. Therefore, eq.\eqref{GW_4} takes the form:
\begin{equation}
    \square e_{\mu\nu}(x) = - T_{\mu\nu}(x).
    \label{GW_5}
\end{equation}

The retarded solution of eq.\eqref{GW_5} is given in eq.\eqref{Rad_fld_8}.
\begin{equation}
    e_{\mu\nu}(x) = -\int d^4 x^\prime \, G_r (x, x^\prime) \, T_{\mu\nu}(x),
    \label{Rad_fld_8}
\end{equation}
where $G_r (x, x^\prime)$ is the retarded Green's function:
\begin{equation}
    G_r (x, x^\prime) = \int \frac{dl^4}{(2 \pi)^4} \frac{e^{i \, l \cdot (x - x^\prime)}}{(l_0 + i \, \epsilon)^2 - \vec{l}^2} \hspace{2mm}, \text{ where } \epsilon \to 0.
    \label{Rad_fld_9}
\end{equation}

As the interest of this paper lies in $D^{(3)}\Xi^\pm = \lim_{u \to \pm\infty} u^2 D^{(3)} N_{\mu\nu}(x)$, this paper intends only to study the linearized perturbation of the Minkowski metric at $\mathcal{I}^+_\pm$. With that goal in mind, the classical soft graviton theorem will be explored, and from there, one can get the asymptotic form of $h_{\mu\nu}$ at $\mathcal{I}^+_\pm$ via inverse Fourier transformation.

\section{Review of Classical Soft Graviton (CSG) Theorem}\label{Sec_CSGT}

In this section, the CSG theorem will be briefly reviewed. Firstly, the CSG theorem will be stated in sec.[\ref{Sec_CSGT_1}]. Then, the algorithm for proving the CSG theorem will be reviewed in sec.[\ref{Sec_CSGT_2}].

\subsection{What does CSG Theorem Say?}\label{Sec_CSGT_1}

The CSG theorem states the behavior of the low-frequency gravitational radiation is of a very particular form. According to the, CSG theorem, different terms in $\Tilde{e}_{\mu\nu}(\omega, \Vec{x})$, at soft limit, will be proportional to $\omega^{(\zeta-1)} (\log{\omega})^\kappa$, where $\zeta, \kappa= \{0,1,2,3,....\}$. Generally, $\Tilde{e}_{\mu\nu}(\omega, \Vec{x})$ at soft limit takes the following form:
\begin{equation}
    \Tilde{e}_{\mu\nu}(\omega, \Vec{x})= \frac{e^{i \omega r}}{4 \pi r} \Big\{\frac{1}{\omega}A_{\mu\nu}(\hat{n}) + \log{\omega} \, B_{\mu\nu}(\hat{n}) + C_{\mu\nu}(\hat{n})+ \mathcal{O}(\omega)\Big\}.
    \label{4.1.1}
\end{equation}
$\mathcal{O}(\omega)$ in eq.\eqref{4.1.1} can be a bit misleading because it includes terms like $\omega^2 \log{\omega}$, $\omega \log{\omega}$, etc.

\subsection{Getting to the CSG Theorem}\label{Sec_CSGT_2}

In this subsection, it will be shown how one can get to the proof of the CSG theorem.

Now, Fourier transforming ${e}_{\mu\nu}(x)$ from eq.\eqref{Rad_fld_8} gives
\begin{align}
    \Tilde{e}_{\mu\nu}(\omega, \vec{x}) \nonumber &=  -\int dt e^{ i \omega t} \int d^4 x^\prime  \int \frac{dl^4}{(2 \pi)^4} \frac{e^{i l \cdot (x - x^\prime)}}{(l_0 + i \epsilon)^2 - \vec{l}^2} T_{\mu\nu}(x^\prime) \\  
    \nonumber&\simeq \frac{e^{i \omega |\Vec{x}|}}{4 \pi |\vec{x}|}\int d^4 x^\prime e^{-i k \cdot x^\prime} T_{\mu\nu}(x^\prime)\\ 
    &= \frac{e^{i \omega r}}{4 \pi r}\int d^4 x^\prime e^{-i k \cdot x^\prime} T_{\mu\nu}(x^\prime),
    \label{Rad_fld_10}
\end{align}

In eq.\eqref{Rad_fld_10}, it has been assumed that $r=|\Vec{x}| >> |\Vec{x^\prime}|$. The detailed calculation \cite{Sen_Laddha} uses saddle point approximation to reach the final result in eq.\eqref{Rad_fld_10}.\\

Upon considering a physical system, one gets the stress-energy tensor $ T_{\mu\nu}$. Then, using that $ T_{\mu\nu}$, one can produce the expression of the form given in eq.\eqref{4.1.1} via eq.\eqref{Rad_fld_10}.

\section{Behavior of the spacetime at $\mathcal{I}^+_\pm$}\label{Sec_Behav}

Now that we know the general statement of the CSG theorem, we can investigate the behavior of the linearized gravitational wave at $\mathcal{I}^+_\pm$. And that will enable us to predict in which cases the peeling will be violated at $\mathcal{I}^+_\pm$ as we have pointed out in sec.[\ref{Sec_Violation of Peeling}]. We will derive the genaralized behavior of $e_{\mu\nu}$ and then, $h_{\mu\nu}$ at $\mathcal{I}^+_\pm$. From the behavior of $h_{\mu\nu}$ at $\mathcal{I}^+_\pm$, we will be able to predict if peeling is violated at $\mathcal{I}^+_\pm$.

\subsection{Behavior of $e_{\mu\nu}(x)$ at $\mathcal{I}^+_\pm$}\label{Sec_Behav_1}

To analyze the behaviour of $e_{\mu\nu}(x)$, we do the inverse Fourier transformation of $\Tilde{e}_{\mu\nu}(\omega, \vec{x})$ to get $e_{\mu\nu}(t, \vec{x})$.
\begin{align}
    e_{\mu\nu}(t, \vec{x})  \noindent\nonumber& =\int \frac{d\omega}{2\pi} e^{-i \omega t}\Tilde{e}_{\mu\nu}(\omega, \vec{x})\\
     \noindent\nonumber& =\int \frac{d\omega}{2\pi} e^{-i \omega t} \frac{e^{i \omega r}}{4 \pi r}\int d^4 x^\prime e^{-i k \cdot x^\prime} T_{\mu\nu}(x^\prime) \\
      \noindent& = \frac{1}{4 \pi r}\int \frac{d\omega}{2\pi}  e^{-i \omega u}\int d^4 x^\prime e^{-i k \cdot x^\prime} T_{\mu\nu}(x^\prime) \hspace{3mm},\hspace{3mm} \text{where } u=t-r.
     \label{GW_7}
\end{align}

We expect $\lim_{\omega \to 0} \Tilde{e}_{\mu\nu}(\omega, \Vec{x})$ to be corresponding to $\lim_{|u| \to \infty} e_{\mu\nu}(t, \vec{x})$, by the principle of Fourier transformation. From the App[\ref{App_Inv_FT}], we can imply that the inverse Fourier transformation of $\Tilde{e}_{\mu\nu}(\omega, \Vec{x})$ at soft limit will give us:

\begin{equation}
    e_{\mu\nu}=   
    \begin{cases}
    \frac{1}{4\pi r}\left(\frac{B_{\mu\nu}(\hat{n})}{u}+\ldots\right)   & \quad \text{at }\mathcal{I}^+_-\\
    \frac{1}{4\pi r}\left(-i A_{\mu\nu}(\hat{n}) -\frac{B_{\mu\nu}(\hat{n})}{u}+\ldots\right)   & \quad \text{at }\mathcal{I}^+_+
  \end{cases}
  \label{5.0.2}
\end{equation}

\subsection{Behavior of $h_{\mu\nu}(x)$ at $\mathcal{I}^+_\pm$}\label{Sec_Behav_2}

Using eq.\eqref{GW_3} and eq.\eqref{5.0.2}, we get:

\begin{equation}
    h_{\mu\nu}=   
    \begin{cases}
    \frac{1}{4\pi r}\left(\frac{1}{u} \left(B_{\mu\nu}(\hat{n}) - \frac{1}{2} \eta_{\mu\nu} \eta^{\alpha\beta}B_{\alpha\beta}(\hat{n})\right) +\ldots\right)   & \quad \text{at }\mathcal{I}^+_-\\
    \frac{1}{4\pi r}\left(-i (A_{\mu\nu}(\hat{n}) - \frac{1}{2} \eta_{\mu\nu} \eta^{\alpha\beta}A_{\alpha\beta}(\hat{n})) -\frac{1}{u} \left(B_{\mu\nu}(\hat{n}) - \frac{1}{2} \eta_{\mu\nu} \eta^{\alpha\beta}B_{\alpha\beta}(\hat{n})\right)+\ldots\right)   & \quad \text{at }\mathcal{I}^+_+
  \end{cases}
  \label{6.0.1}
\end{equation}

\section{Effect of Soft Gravitational Radiation on Future Null Infinity}\label{Sec_Eff_Sft_Grav}

In this section, it is stated how different terms in the soft expansion of gravitational radiation given in eq.\eqref{4.1.1} influence the structure of $\mathcal{I}^+$.

\subsection{Effect of $\frac{1}{\omega}$ Terms at Soft Limit}\label{Sec_Eff_Sft_Grav_1}

Following eq.\eqref{L_E_Fourier_17},eq.\eqref{5.0.2} and eq.\eqref{6.0.1}, we can conclude that if there is a term that is proportional to $\frac{1}{\omega}$ in $\tilde{e}_{\mu\nu} (\omega, \Vec{x})$ at the soft limit, there will be a term proportional to $H(u)$ in $e_{\mu\nu} (t, \Vec{x})$ at $\mathcal{I}^+_\pm$. Here, $H(u)$ is the Heavyside theta function. Because of the presence of this step function in $e_{\mu\nu} (t, \Vec{x})$, there is a jump in $h_{\mu\nu} (t, \Vec{x})$ at $\mathcal{I}^+$ as we go from $u \to -\infty$ to $u \to \infty$. This we can see in eq.\eqref{5.0.2}. 

From this, we can see how $\frac{1}{\omega}$ term in $\tilde{e}_{\mu\nu} (\omega, \Vec{x})$ at the soft limit is responsible for the gravitational memory effect at $\mathcal{I}^+$. But for the $\frac{1}{\omega}$ term in $\tilde{e}_{\mu\nu} (\omega, \Vec{x})$ at the soft limit, the contribution in $D^{(3)}\Xi^{\pm}$ is $0$. So, the $\frac{1}{\omega}$ term in $\tilde{e}_{\mu\nu} (\omega, \Vec{x})$ does not violate the peeling.

\subsection{Effect of $\log{\omega}$ Terms at Soft Limit}\label{Sec_Eff_Sft_Grav_2}

If we compare eq.\eqref{L_E_Fourier_22}, eq.\eqref{5.0.2} and eq.\eqref{6.0.1}, we can conclude that if there is a term that is proportional to $\log{\omega}$ in $\tilde{e}_{\mu\nu} (\omega, \Vec{x})$ at the soft limit, there will be a term proportional to $\frac{1}{u}$ in $h_{\mu\nu} (t, \Vec{x})$. This can lead to a finite and non-zero value of $D^{(3)}\Xi^{\pm}$.

Therefore, $\log{\omega} $ term in $\tilde{e}_{\mu\nu} (\omega, \Vec{x})$ at soft limit can responsible for violation of peeling at $\mathcal{I}^+$ as $u \to \pm \infty$.

\subsection{Effect of Other Terms at Soft Limit}\label{Sec_Eff_Sft_Grav_3}

In the soft limit, the other terms in $\tilde{e}_{\mu\nu} (\omega, \Vec{x})$ do not contribute to $D^{(3)}\Xi^{\pm}$. Hence, they are not responsible for the violation of peeling.

\section{Results and Discussion}\label{Sec_Rslt_Disc}

 It is shown in sec.[\ref{Sec_Eff_Sft_Grav_1}] that the $\frac{1}{\omega}$ terms in $\lim_{\omega \to 0}\Tilde{e}_{\mu\nu} (\omega, \Vec{x})$ are responsible for the memory effect at $\mathcal{I}^+$. It is also demonstrated in sec.[\ref{Sec_Eff_Sft_Grav_2}] that the $\log{\omega}$ terms in $\lim_{\omega \to 0}\Tilde{e}_{\mu\nu} (\omega, \Vec{x})$ generate the terms proportional to $\frac{1}{u}$ in $\lim_{r, |u| \to \infty}h_{\mu\nu}(x)$ at leading order. Sec.[\ref{Sec_Violation of Peeling}] shows how the terms proportional to $\frac{1}{u}$ in $\lim_{r, |u| \to \infty}h_{\mu\nu}(x)$ at leading order generate non-artifact $\log{r}$ terms and make the future null infinity $C^{<3}$ if $\mathcal{D}^{(3)}\Xi^\pm$ is non-zero and finite. Therefore, the $\log{\omega}$ terms $\lim_{\omega \to 0}\tilde {e}_{\mu\nu} (\omega, \Vec{x})$ are responsible for the violation of peeling at $\mathcal{I}^+$ as $u \to \pm \infty$ if the $\log$ term has appropriate coefficients such that $\mathcal{D}^{(3)}\Xi^\pm$ is non-zero and finite. Hence, for a system, if there is term proportional $\log{\omega}$ present in $\lim_{\omega \to 0}\tilde {e}_{\mu\nu} (\omega, \Vec{x})$ such that $\mathcal{D}^{(3)}\Xi^\pm$ is non-zero and finite, peeling will be violated at $\mathcal{I}^+_\pm$.




\vspace{5mm}

\hrule

\begin{center}
    \huge Appendix
\end{center}

\appendix
\section{Inverse Fourier Transformation of Few Typical Terms in Soft Expansions}\label{App_Inv_FT}

In the analysis of soft radiation, it is common to deal with functions $\Tilde{F}(\omega, \Vec{x})$ \cite{Sen}, that are non-analytic as $\omega \to 0$. Different terms in $\Tilde{F}(\omega, \Vec{x})$ will be proportional to $\omega^{(\zeta-1)} (\log{\omega})^\kappa$, where $\zeta, \kappa= \{0,1,2,3,....\}$. It is expected that $\lim_{\omega \to 0} \Tilde{F}(\omega, \Vec{x})$ to be corresponding to $\lim_{|u| \to \infty} F(t, \vec{x})$, by the principle of Fourier transformation. Now, let's derive the precise correspondance between $\lim_{\omega \to 0} \Tilde{F}(\omega, \Vec{x})$ and $\lim_{|u| \to \infty} F(t, \vec{x})$. All this analysis is done for constant $\Vec{x}$. So, we will not display the $\Vec{x}$ dependence.

\subsection{\underline{Case 1: $\zeta=\kappa=0$}}\label{Sec_L_E_Fourier_Case_1}

The first thing done here is examining small $\omega$ singularities of the form $\frac{1}{\omega}$. Let's assume that the function is of the form: $\Tilde{F}(\omega) = C \, e^{i \omega \phi} \frac{1}{\omega}f(\omega)$, where $C$ and $\phi$ is constant, and $f(\omega)$ is a function of $\omega$, such that $f(\omega)$ is smooth at $\omega=0$ $f(0)=1$.   

\begin{equation}
    F(t) = \int \frac{d\omega}{2 \pi} e^{-i \omega t} \Tilde{F}(\omega) = C \int \frac{d\omega}{2 \pi} e^{- i \omega u} \frac{1}{\omega} f(\omega)\hspace{3mm},\hspace{3mm} \text{where } u\equiv t-\phi.
    \label{L_E_Fourier_12}
\end{equation}

Let's assume,
\begin{equation}
     F_\pm (t) = \frac{C}{2 \pi} \int d\omega e^{-i \omega u} \frac{1}{\omega \pm i \epsilon} f(\omega).
     \label{L_E_Fourier_14}
\end{equation}

From eq.\eqref{L_E_Fourier_14}, one gets:
\begin{align}
    F_-(t) - F_+(t) \nonumber\noindent& = \frac{C}{2 \pi} \int d\omega e^{-i \omega u} \frac{1}{\omega - i \epsilon} f(\omega) - \frac{C}{2 \pi} \int d\omega e^{-i \omega u} \frac{1}{\omega + i \epsilon} f(\omega) \\
    \noindent\nonumber& =   i \, C \int d\omega e^{-i \omega u} \delta(\omega) f(\omega)  \\
    \noindent& =   i \, C  ,
    \label{L_E_Fourier_15}
\end{align}
where the following relation has been used: $\delta(\omega) = \frac{1}{2\, \pi \, i}\lim_{\epsilon \to 0} (\frac{1}{\omega-i\epsilon} - \frac{1}{\omega+i\epsilon}). $

$F_-(t) - F_+(t)$ is just a constant. Hence, $F_-(t)$ and $F_+(t)$ are only differ by a quantity independent of $u$. The difference is of no interest to this article. For definiteness, one can only work with $F_+(t)$.

To do the integration in eq.\eqref{L_E_Fourier_14} (for $F_+(t)$ case), one can close the contour in the lower (upper) half plane for positive (negative) $u$ and pick up the residues at the poles. 
  
Now, one gets:
\begin{eqnarray}
     \noindent\nonumber&& F_+(t) = \frac{C}{2 \pi} \int d\omega \, e^{-i \omega u} \frac{1}{\omega + i \epsilon} f(\omega)=
  \begin{cases}
    -i \, C   + \mathcal{O}(e^{-u})    & \quad \text{for } u \to \infty\\
    \mathcal{O}(e^{-u})   & \quad \text{for } u \to -\infty
  \end{cases}
  \label{L_E_Fourier_16}
\end{eqnarray}

The result in eq.\eqref{L_E_Fourier_16} can be summarized as following:
\begin{equation}
    F_\pm (t) = -i \, C \, H(u) + \mathcal{O}(e^{-u})
    \label{L_E_Fourier_17}
\end{equation}

$H(u)$ in eq.\eqref{L_E_Fourier_15} is the Heavyside Theta function defined as:

$$  H(u) =
  \begin{cases}
    1       & \quad \text{if } u \geq 0\\
    0   & \quad \text{if } u <0
  \end{cases} $$

The $\mathcal{O}(e^{-u})$ contribution in eq.[\eqref{L_E_Fourier_16}] comes from the poles of $f(\omega)$.

\subsection{\underline{Case 2: $\zeta=\kappa=1$}}\label{Sec_L_E_Fourier_Case_2}

Now, let's analyze the singularities of the form $\log{\omega}$ for small $\omega$. Let's assume that the function is of the form: $\Tilde{F}(\omega) = C e^{i \omega \phi} (\log{\omega}) f(\omega)$, where $C$ and $\phi$ is constant, and $f(\omega)$ is a function of $\omega$, such that $f(\omega)$ is smooth at $\omega=0$ $f(0)=1$.

Now, let's define:
\begin{equation}
     F_\pm (t) = \frac{C}{2 \pi} \int d\omega \, e^{-i \omega u} \log{(\omega \pm i \epsilon)} f(\omega).
     \label{L_E_Fourier_18}
\end{equation}

Therefore,
\begin{equation}
      F_-(t) - F_+(t)  = \frac{C}{2 \pi} \int d\omega \, e^{-i \omega u} \log{(\omega - i \, \epsilon)} f(\omega) - \frac{C}{2 \pi} \int d\omega \, e^{-i \omega u} \log{(\omega + i \,  \epsilon)} f(\omega)
      \label{L_E_Fourier_19}
\end{equation}

To do the calculation in eq.\eqref{L_E_Fourier_19}, both sides of the expression are integrated: $\delta(\omega) = \frac{1}{2\, \pi \, i}\lim_{\epsilon \to 0} (\frac{1}{\omega-i\epsilon} - \frac{1}{\omega+i\epsilon})$ with respect to $\omega$.

\begin{align}
     \noindent\nonumber& \hspace{8mm}\int d\omega \, \delta(\omega) = \frac{1}{2\, \pi \, i}\lim_{\epsilon \to 0} \int d\omega \,   \Big{(}\frac{1}{\omega-i\epsilon} - \frac{1}{\omega+i\epsilon}\Big{)} \\
     \noindent& \implies \lim_{\epsilon \to \,  0} \big( \log{(\omega + i \epsilon)}) -(\log{(\omega - i \epsilon)} \big) = 2\pi i H(-\omega) . 
    \label{L_E_Fourier_20}
\end{align}

From eq.\eqref{L_E_Fourier_19} and eq.\eqref{L_E_Fourier_20}, one gets:
\begin{align}
       F_+(t) - F_-(t) \noindent\nonumber&  = i \, C \int d\omega \, e^{-i \omega u} H(-\omega) f(\omega) \\
        \noindent\nonumber& = i \, C \int_{-\infty}^0 d\omega \,  e^{-i \omega u} f(\omega) \\
        \noindent& \simeq -\frac{C}{u} \hspace{3mm},\hspace{3mm} \text{for } |u| \to \infty.
      \label{L_E_Fourier_21}
\end{align}
The $\mathcal{O}(e^{-u})$ terms, that arise due to the poles of $f(\omega)$, have been suppresed.

Therefore,
\begin{eqnarray}
     \noindent\nonumber&& F_+(t) = \frac{C}{2 \pi} \int d\omega \, e^{-i \omega u} \log{(\omega + i \, \epsilon)} \, f(\omega)=
  \begin{cases}
    -\frac{C}{u}    & \quad \text{for } u \to \infty\\
    0   & \quad \text{for } u \to -\infty
  \end{cases} \\
    \noindent&& F_-(t) = \frac{C}{2 \pi} \int d\omega \, e^{-i \omega u} \log{(\omega - i \,  \epsilon)} \, f(\omega)=
  \begin{cases}
    0    & \quad \text{for } u \to \infty\\
    \frac{C}{u}    & \quad \text{for } u \to -\infty
  \end{cases} 
  \label{L_E_Fourier_22}
\end{eqnarray}

\subsection{\underline{Case 3: $\zeta=2$ and $\kappa=1$}}\label{Sec_L_E_Fourier_Case_3}

Now, let's analyze the singularities of the form $\log{\omega}$ for small $\omega$. Let us assume the function of the form: $\Tilde{F}(\omega) = C e^{i \omega \phi}\omega (\log{\omega}) f(\omega)$,  where $C$ and $\phi$ is constant, and $f(\omega)$ is an function of $\omega$, such that $f(\omega)$ is smooth at $\omega=0$ $f(0)=1$.

Now, let's define:
\begin{align}
     F_\pm (t) \noindent\nonumber&  = \frac{C}{2 \pi} \int d\omega \, e^{-i \omega u} \omega\log{(\omega \pm i \, \epsilon)} \, f(\omega) \\
     \noindent& = \frac{i \, C}{2 \pi} \frac{d}{du} \int d\omega \, e^{-i \omega u} \log{(\omega \pm i \, \epsilon)} \, f(\omega).
     \label{L_E_Fourier_23}
\end{align}

Using eq.\eqref{L_E_Fourier_22} and eq.\eqref{L_E_Fourier_23}, one gets:
\begin{eqnarray}
     \noindent\nonumber&& F_+(t) = \frac{C}{2 \pi} \int d\omega \, e^{-i \omega u} \omega\log{(\omega + i \, \epsilon)} \, f(\omega)=
  \begin{cases}
    i\frac{C}{u^2}    & \quad \text{for } u \to \infty\\
    0   & \quad \text{for } u \to -\infty
  \end{cases} \\
    \noindent&& F_-(t) = \frac{C}{2 \pi} \int d\omega \, e^{-i \omega u} \omega\log{(\omega - i \, \epsilon)} \, f(\omega)=
  \begin{cases}
    0    & \quad \text{for } u \to \infty\\
    -i\frac{C}{u^2}    & \quad \text{for } u \to -\infty
  \end{cases} 
  \label{L_E_Fourier_24}
\end{eqnarray}

\subsection{\underline{Case 4: $\zeta=\kappa=2$}}\label{Sec_L_E_Fourier_Case_4}

Now, let's analyze the singularities of the form $\log{\omega}$ for small $\omega$. Let's assume the function of the form: $\Tilde{F}(\omega) = C e^{i \omega \phi}\omega \{(\log{\omega})\}^2 f(\omega)$, where $C$ and $\phi$ is constant, and $f(\omega)$ is an function of $\omega$, such that $f(\omega)$ is smooth at $\omega=0$ $f(0)=1$.

Now, let's define:
\begin{align}
     F_\pm (t) \noindent\nonumber&  = \frac{C}{2 \pi} \int d\omega \, e^{-i \omega u} \omega \, \{\log{(\omega \pm i \, \epsilon)} \}^2 \, f(\omega) \\
     \noindent& = \frac{i \, C}{2 \pi} \frac{d}{du} \int d\omega \, e^{-i \omega u} \{\log{(\omega \pm i \, \epsilon)}\}^2 \,  f(\omega)
     \label{L_E_Fourier_25}
\end{align}\\
and
\begin{equation}
     G (t) = \frac{C}{2 \pi} \int d\omega \, e^{-i \omega u} \omega\{\log{(\omega + i \,  \epsilon)}\} \, \{\log{(\omega - i \,  \epsilon)}\} \, f(\omega). 
     \label{L_E_Fourier_26}
\end{equation}

\begin{eqnarray}
     \noindent\nonumber&& F_+(t) = \frac{C}{2 \pi} \int d\omega \, e^{-i \omega u} \omega\{\log{(\omega + i \,  \epsilon)}\}^2 \, f(\omega)=
  \begin{cases}
    -2 \, i \, C\frac{\log{|u|}}{u^2}    & \quad \text{for } u \to \infty\\
    0   & \quad \text{for } u \to -\infty
  \end{cases} \\
    \noindent&& F_-(t) = \frac{C}{2 \pi} \int d\omega \, e^{-i \omega u} \omega\log{(\omega - i \, \epsilon)} \, f(\omega)=
  \begin{cases}
    0    & \quad \text{for } u \to \infty\\
    2 \, i \, C\frac{\log{|u|}}{u^2}    & \quad \text{for } u \to -\infty
  \end{cases} 
  \label{L_E_Fourier_27}
\end{eqnarray}

\begin{equation}
     G (t) = \frac{C}{2 \pi} \int d\omega \, e^{-i \omega u} \omega\{\log{(\omega + i \,  \epsilon)}\} \, \{\log{(\omega - i \epsilon)}\} \, f(\omega)=
     \begin{cases}
    -i \, C \, \frac{\log{|u|}}{u^2}    & \quad \text{for } u \to \infty\\
    i \, C \, \frac{\log{|u|}}{u^2}   & \quad \text{for } u \to -\infty
  \end{cases}. 
     \label{L_E_Fourier_28}
\end{equation}

\subsection{\underline{Case 5: $\zeta=1,2,3,...$ and $\kappa=0$}}\label{Sec_L_E_Fourier_Case_5}

Now, let's analyze ${\omega^n}$ for small $\omega$, where $n=0,1,2,....$. Let's assume the function of the form: $\Tilde{F}(\omega) = C e^{i \omega \phi}\omega^n f(\omega)$, where $C$ and $\phi$ is constant, and $f(\omega)$ is an function of $\omega$, such that $f(\omega)$ is smooth at $\omega=0$ $f(0)=1$.

Now, one defines:
\begin{align}
    F(t) \noindent\nonumber& =\frac{C}{2\pi} \int d\omega \, e^{-i \omega u} \, \omega^n \, f(\omega) \\
    \noindent\nonumber& =\frac{C}{2\pi} {(-i)}^n \frac{d^n}{du^n}\int d\omega \, e^{-i \omega u} f(\omega) \\
     \noindent& ={(-i)}^n \frac{d^n}{du^n} \{\mathcal{O}(e^{-u})\}
\end{align}

\bibliographystyle{unsrt}  

\end{document}